# Multipass Faraday rotators and isolators

**Johann Gabriel Meyer*, Andrea Zablah, Kristaps Kapzems, Nazar Kovalenko, and Oleg Pronin**

*Helmut Schmidt University, Holstenhofweg 85, 22043 Hamburg, Germany*
**johann.meyer@hsu-hh.de*

**Abstract:** Faraday isolators are usually limited to Faraday materials with strong Verdet constants. We present a method to reach the 45° polarization rotation angle needed for optical isolators with materials exhibiting a weak Faraday effect. The Faraday effect is enhanced by passing the incident radiation multiple times through the Faraday medium while the rotation angle accumulates after each pass. Materials having excellent thermos-optical properties in the ultraviolet and mid-infrared range become available for optical isolators. Herriott-type multipass cells offer a simple and compact way to realize the desired propagation length in usual optical materials of standard sizes. A proof-of-principle experiment was carried out, demonstrating polarization rotation of a 532 nm laser beam by an angle of 45° in anti-reflection-coated fused silica surrounded by a standard neodymium ring magnet.

## 1. Introduction

Optical isolators are a crucial component where laser setups are sensitive to back reflections of optical radiation. Usually, optical isolation is realized with Faraday isolators. Between two polarizers rotated by 45° against each other, the polarization of the incident radiation is rotated employing the Faraday effect within a Faraday medium [Fig. 1]. A magnetic field is applied to the medium, whereby the field is aligned parallel or antiparallel to the propagation direction of the radiation. For a beam entering the isolator from one side, the resulting Faraday effect rotates the polarization by an angle of 45° so that full transmission through the second polarizer (pol. 2) is granted. When the exiting beam is reflected back into the isolator, its polarization is also rotated by an angle of 45°. Viewed from the fixed laboratory frame, the polarization rotation for both passes adds up, resulting in a polarization oriented 90° towards the polarizer 1, which leads to a total rejection of the reflected beam. The rotation angle for a single pass is given by:

$$\varphi = V \int_0^l B_z(z) dz$$

The rotation angle is determined by the medium's Verdet constant *V*, the component of the magnetic field parallel to the propagation direction of the optical radiation $B_z$, and the propagation length within the material *l*. When the propagation direction is reversed towards the magnetic field vector, the integral requires a change in the sign. Therefore, the direction of the polarization rotation reverses with regard to the wave vector. However, when observed from the fixed laboratory frame, the Faraday rotator rotates the electric field vector of the passing electromagnetic wave always in the same direction. Therefore, the rotation angle of the electric field vector accumulates when an electromagnetic wave passes through a Faraday medium a second time after being reflected back into it by a reflecting surface. It is important to note that the reflecting surface changes the direction of propagation but not the polarization direction.

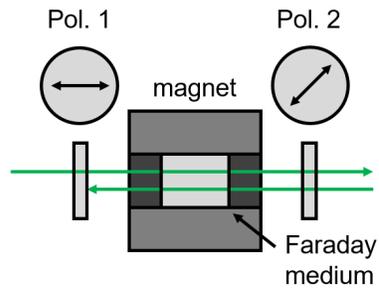

**Figure 1.** Schematic of conventional Faraday isolator. The first polarizer Pol. 1 ensures linear polarization for the entering beam. An analyzing polarizer Pol. 2 is rotated by an angle of 45° with respect to the first polarizer. The Faraday medium is placed in a magnetic system, here a simple ring magnet, to get a magnetic field collinear with the beam propagation direction.

In Faraday isolators for near-infrared (NIR) lasers with high average powers, terbium gallium garnet (TGG) is a commonly used Faraday medium [1,2]. It offers an exceptional Verdet constant of about 39 rad/(T·m) at a wavelength of 1 µm and an absorption coefficient of about $1 \cdot 10^{-3}$ cm$^{-1}$[2]. Another upcoming Faraday medium is potassium terbium fluoride (KTF) with a similar Verdet constant and an order of magnitude lower absorption and thermo-optic coefficients, allowing for the handling of higher average powers. KTF, however, is more difficult to grow [1]. Typically, a rod of TGG with a length of a few mm is placed inside a system of permanent ring magnets providing the axial magnetic field on the order of magnitude of 1 T [Fig. 1]. Practically, the operation is limited to average powers of a few 100 W [2]. The limiting factor here is the degrading isolation performance due to temperature gradients from the absorption of optical radiation. Whereas thermal lensing and temperature-dependent shifts in the Verdet constant are problematic, the main limitation arises from birefringence induced by the photo-elastic effect under thermal load [3]. Intricate schemes for compensation of the birefringence within the isolator allow for average powers beyond 1 kW while maintaining an isolation performance of 30 dB [2,4]. Another approach has been proposed [5], in which the thermal gradients inside the Faraday medium are reduced by a cooling geometry known for its effective application in high-power thin-disk lasers [6]. Thereby, the Faraday medium is prepared in the form of a thin-disk and mounted from one side onto a water-cooled heatsink. The reduced thickness of the thin-disk results in a reduced Faraday rotation, which has to be compensated by several passes over the disk.

Generally, media with a low amount of absorption, thermal expansion, and thermo-optic coefficient are desired to ensure low depolarization from temperature gradients. At a wavelength of 1 µm, quartz and fused silica have an exceptionally low absorption coefficient, up to two orders of magnitude lower than TGG, and also a slightly lower thermo-optic coefficient [Tab. 1]. However, due to the low Verdet constant, it becomes necessary to increase the propagation length inside the material [7], making it impractical to use in typical Faraday isolator schemes [Fig. 1]. Whereas, already at a wavelength of 1 µm, the low absorption might balance the increased propagation length in terms of power limitation due to depolarization. In the visible to ultraviolet (UV) range, TGG suffers increasingly from absorption due to color centers and, thus, from enhanced thermal degradation [1]. Other materials are required there. Faraday isolators at 405 nm and 355 nm have already been demonstrated based on cerium(III) fluoride (CeF$_3$) [8], which enables transmission down to 300 nm [9]. For the UV range down to 200 nm, PrF$_3$ was presented as a potential Faraday medium with a high Verdet constant [9]. On the other hand, materials like quartz and magnesium fluoride (MgF$_2$) are readily available and offer a similar transmission range down to about 200 nm. As the latter two materials suffer from lower Verdet constants, they may be used as a Faraday medium inside a Faraday isolator employing multiple passes through the medium. Additionally, MgF$_2$ exhibits an order of magnitude lower thermo-optic coefficient than TGG and quartz, which could reduce thermal effects [10,11].

## 2. Concept

In the following, we will discuss briefly the obstacles when employing Faraday media with low Verdet constants and then present a multipass approach enabling sufficient polarization rotation angles for Faraday isolators with these materials. In the end, we will also discuss the applicability of this approach to gases as Faraday media, which could be beneficial when operating higher average powers, even though gases suffer from much lower Verdet constants [12,13].

**Table 1. Material properties of different Faraday media**

| Properties | TGG | Quartz/fused silica | MgF$_2$ |
|---|---|---|---|
| Verdet constant (rad/(T·m)) | 39 [1] (1064 nm) | 1.1 [14,15] (1064 nm, silica fiber) 5.5 [16] (515 nm, fused silica) 7.0 [16] (458 nm, fused silica) | 2.4 [17] (633 nm) 4.6 [17] (458 nm) |
| Absorption coefficient (10$^{-3}$ cm$^{-1}$) | 0.7 – 1.5 [2] (1 µm) | ca. 0.02 [18] (1 µm, fused silica) | <1 / 10 [11] (o./un-pol., 290 nm)[a] <1 / 26 [11] (o./un-pol., 220 nm)[a] 13 / 56 [11] (o./un-pol., 200 nm)[a] |
| Thermal conductivity at 300 K (W/(m·K)) | 4.5 [19] | 6.2 / 10.4 [10] (a/c-axis) | 30 / 21 [10] (a/c-axis) |
| Thermal expansion at 300 K (ppm/K) | 7.3 [20] | 12.4 / 6.9 [10] (a/c-axis) | 9.4 / 13.6 [10] (a/c-axis) |
| Thermo-optic coefficient at 300 K (ppm/K) | 17.9 [20] (633 nm) | ca. -8 [21] (1 µm) -6.2 / 7.0 [10] (546 nm) | 1.1/0.6 [10] (o/e-ray., 633 nm)[b] 0.9 [10] (458 nm) |

[a]) polarization normal to the optical axis, i.e., ordinarily ray (o.) or unpolarized (un-pol.)
[b]) ordinary (o) or extraordinary (e) ray

To achieve a polarization rotation on the order of 45° in materials with low Verdet constants, either magnetic fields with extraordinarily high flux densities must be applied to the material or the propagation length in the material must be increased drastically. The magnetic flux density provided by permanent neodymium magnets is typically on the order of magnitude of 1 T. Taking this as a practical limitation for the magnetic field, increasing the propagation lengths to tens of cm, if not several meters, becomes necessary.

Another complication arises from the necessity to apply a magnetic field over such a long path when permanent magnets are used. In the simplest case, the Faraday medium is surrounded by a ring magnet. Increasing just the length of the ring magnet, along with the length of the medium, leads to an almost vanishing field inside the magnet, which is due to the inversion of the field direction in the far field of a magnetic dipole [Fig. 4]. To keep a reasonable field strength, the outer diameter of the ring magnet must be scaled by the same order of magnitude as its length. Permanent magnets with the resulting dimensions tend to be impractical. On the other hand, electromagnets, i.e., solenoids, can, in principle, be scaled to the required lengths without sacrificing any magnetic field inside [7]. But driving currents of several amperes would have to be supplied constantly. Unless sufficient cooling or even superconductors are applied, the magnetic field strength would be typically below that obtained with permanent magnets. In addition, current fluctuations are directly translated into fluctuations of the resulting polarization rotation. A polarizer, as required in Faraday rotators, will translate these fluctuations into amplitude noise. Furthermore, polarization rotation fluctuations will lead to a degraded isolation performance in Faraday isolators. This makes it necessary to supply a sufficiently stable current when the setup is sensitive to noise or small back reflections.

For media with low Verdet constants, the necessarily longer path length can be realized with optical fibers when considering glasses like fused silica. Faraday rotation of several degrees has been demonstrated inside a fused silica fiber rolled up multiple times through the magnetic field of solenoids [14]. Thereby, bend-induced birefringence in the optical fiber needs to be compensated to ensure a constructive buildup of Faraday rotation after the polarization flips due to the birefringence [14,22]. Furthermore, a small amount of uncompensated birefringence generally leads to an elliptical polarization, which limits the isolation performance of the Faraday isolator. However, coupling into single-mode fibers makes this approach challenging to use with powerful free-space laser beams.

Therefore, we present here an approach based on free-space optics to produce enhanced Faraday rotation in media with low Verdet constants for potential use in Faraday isolators. Similar to Faraday rotation spectroscopy setups for high sensitivity [23–25], the Faraday medium is placed inside a multipass cell [Fig. 2]. The effective path length inside the medium is increased to the desired amount by guiding the laser beam several times through the employed medium. In the case of solid media, the length of the medium could be approximately a cm or a few mm. For gases as Faraday media, longer distances or more passes would be necessary. The desired amount of passes can be achieved with low complexity in a Herriott cell [24,26]. The multipass approach does not only reduce the requirement for the length of the medium, but it also limits, in the same way, the distance over which a magnetic field has to be applied. Therefore, standard-size systems of permanent ring magnets can be employed.

## 3. Experimental

As a proof of concept, we demonstrated Faraday rotation by an angle of 45° inside a Faraday medium with a low Verdet constant using a continuous-wave laser beam (Coherent Verdi V10) with an output power below 8 W and a wavelength of 532 nm. We chose fused silica for the Faraday medium. It is well suited for high-power laser beams due to its low absorption and weak thermal effects. Its amorphous structure leads to the absence of birefringence.

Two anti-reflection-coated fused silica plates with a length of 6.35 mm each and a diameter of 12.7 mm were mounted inside a ring magnet, providing a mostly uniform magnetic field pointing in the direction of the surface normal of the plates [Fig. 2]. Special care was taken during mounting to avoid any stress-induced birefringence inside the plates. Even the slightest pressure on the plate can lead to a significant disturbance of the resulting rotation. A ring magnet was placed around the fused silica. The magnet has a length of 30 mm, an inner diameter of 22 mm, and an outer diameter of 76 mm. It is a neodymium magnet of the grade N35SH, corresponding to a remanent magnetization of approximately 1.2 T. Based on that, the maximum flux density inside the hollow core of the ring magnet was calculated to be about 0.53 T [Fig. 3] [27]. The Herriott-type multipass cell was built around the fused silica to propagate the laser beam multiple times through it. The number of passes can be easily adjusted by the alignment of the cell. The cell consists of two 1-inch diameter concave spherical mirrors, $CM_1$ and $CM_2$, with a radius of curvature (ROC) of 150 mm and 100 mm, respectively.

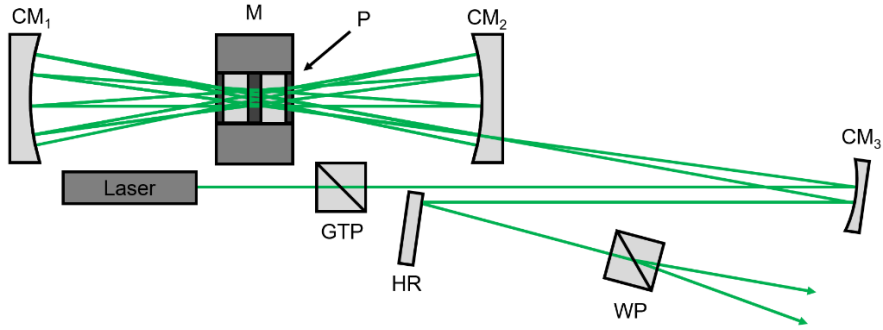

**Figure 2.** Schematic of the Faraday-rotator based on a Herriott cell. $CM_1$: curved mirror (ROC = 150 mm), $CM_2$: curved mirror (ROC = 100 mm), $CM_3$: curved mirror (ROC = 600 mm), M: ring magnet, P: AR-coated plates (fused silica) placed inside the ring magnet.

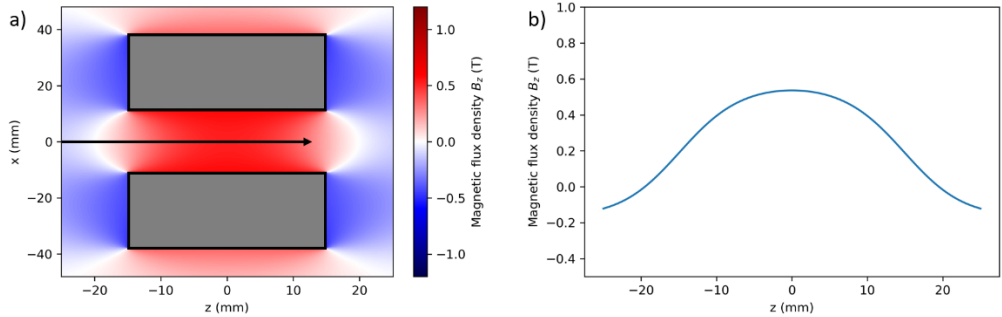

**Figure 3.** a) Simulation with magpylib of the z-component of the magnetic field inside the ring magnet (grey) which is used in the experimental setup [27]. The dimensions are: outer diameter of 76 mm, inner diameter of 22 mm, and a length of 30 mm. Inside the magnet, a homogeneous remanent magnetization of 1.2 T was assumed. The beam propagation direction along the z-axis is indicated by a black arrow. (b) Simulated z-component of the magnetic field on the z-axis going through the center of the ring magnet.

A linearly polarized $TEM_{00}$ laser beam with a wavelength of 532 nm (Coherent Verdi V10) was sent into the setup. A power of up to 8 W was measured right before the setup. The polarization of the input beam was cleaned up by an uncoated Glan-Taylor polarizer (Thorlabs GT10, see GTP in Fig. 2) with an extinction ratio of $10^{-5}$ specified for the transmitted beam. The laser beam was coupled into the cell via a hole in the mirror $CM_2$. The beam was focused into the cell by another concave spherical mirror $CM_3$ with a ROC of 600 mm. The ROC values of the mirrors were chosen to roughly match the eigenmode of the cell to the caustic of the incident laser beam. The beam exiting the cell was collimated by the mirror $CM_3$. Incoming and exiting beams were reflected on that mirror with a separation of a few mm.

The polarization of the exiting beam was checked with an analyzing polarizer, which can be rotated around the propagation direction of the beam. Here, a Wollaston polarizer (Thorlabs WPA10, see WP in Fig. 2) was employed to split the radiation into two beams with orthogonal polarization. For this type of polarizer, an extinction ratio of $10^{-5}$ is specified for both output beams, which allows for a precise investigation of the polarization state. To measure the rotated polarization after the cell, the analyzing polarizer was rotated around the beam axis to minimize the power of one of the output beams of the polarizer. For 30 passes through the fused silica plates, a polarization rotation by an angle of 45° was achieved. The angle was fine-tuned by shifting the fused silica plates inside the magnet. The plates were mounted to be centered on the central axis of the ring magnet and could be moved along this axis. The variation of the z-component of the magnetic field along the z-direction led to a variation of the rotation angle (Fig. 2 (b)). The Faraday effect was verified by reversing the magnetic field direction. The rotation angle reversed to 45° in the opposite direction. This clearly shows that the rotation is not due to mirror reflections or birefringence. It is originating in the accumulation of the Faraday effect inside the Herriott cell. Principally, an empty Herriott cell could rotate the polarization in one direction for specific configurations. We evaluated this potential offset by measuring the polarization rotation angle without the magnet. The resulting angle for our cell configuration was about 0° with a tolerance of 1°. Generally, an offset angle inflicted by the cell itself would not affect the application as a Faraday isolator, as this rotation is reversed for the reflected beam due to its geometrical origin. The effective path length inside the fused silica adds up to 381 mm. The observed rotation angle is in good agreement with the theoretical prediction, considering the literature value for the Verdet constant of fused silica of about 5 rad/(T·m) at a wavelength of 532 nm [15,16].

For a high isolation performance of the Faraday isolator, the polarization of the rotated beam must stay perfectly linear. Otherwise, the isolation performance is reduced. We investigated the degree of linear polarization after the Faraday rotator by comparing the optical power of the two orthogonally polarized beams of the analyzing Wollaston polarizer. The depolarization is characterized by the ratio $\gamma$ of the depolarized power, which is contained in the minimized beam of the analyzing polarizer, to the total power of both beams from the analyzing polarizer:

$$\gamma = \frac{P_{\text{depol}}}{P_{\text{tot}}}$$

For a polarization rotation of 45°, the depolarization ratio assumed during the initial alignment of the setup commonly values of approximately $1 - 2 \cdot 10^{-2}$. The depolarized beam contained a power of about 50 – 100 mW, while the other contained 4 – 5 W. It was possible to reduce the depolarization even further with careful alignment of the beam path inside the Herriott cell and by fine-tuning the position of the arrangement of fused silica plates together with the magnet in relation to the beams. During this alignment procedure, we made sure to maintain the rotation angle of 45°. We could repetitively achieve depolarization values below $2 \cdot 10^{-3}$. We measured, in one case, a power of 8 mW in the depolarized and 4.6 W in the main beam. In an extreme case, it was possible to bring the depolarization down to $2.5 \cdot 10^{-4}$ with 1 mW in the depolarized beam and 4.1 W in the main beam. The lowest values of depolarization were usually obtained when the beams of the Herriott cell were passing the fused silica plates close to the edge. This was achieved by shifting the magnet together with the plates transversally in relation to the beams. Furthermore, the beam path inside the Herriott cell could be aligned to produce an elliptical pattern of reflections on the mirrors to bring all beams closer to one edge of the plates.

A background value for the depolarization ratio was measured in the absence of the Faraday effect. Therefore, we removed the magnet and placed only the fused silica plates inside the Herriott cell. The depolarization ratio assumed a value of $1.5 \cdot 10^{-4}$ with a power of 0.7 mW in the depolarized and 4.8 W in the main beam.

The setup showed 2 – 3 W of losses in total power mostly due to Fresnel reflection on the uncoated polarizers and due to imperfections of the dielectric coatings on the mirrors and fused silica plates. The Herriott cell including the fused silica plates transmitted about 85 % of the incoming radiation. With specialized and optimized coatings for 532 nm wavelength, it would be possible to reach a transmission over 90% [28].

## 4. Discussion

The results demonstrate clearly how the Faraday rotation inside media with low Verdet constants like fused silica can be easily scaled up by using Herriott cells. Employing two simple anti-reflection-coated plates of fused silica with a combined thickness of 12.7 mm resulted in a rotation angle of 45° at a wavelength of 532 nm. Hereby, the angle of 45°, required for Faraday isolators, has been successfully demonstrated. Commercial Faraday isolators typically achieve isolation ratios of about 30 dB. In our setup, we reproducibly achieved depolarization ratios of about $2 \cdot 10^{-3}$, corresponding to an isolation ratio of 27 dB. Without careful alignment, the value could decrease to isolation ratios below 20 dB. The origin of this variation is not completely clear. It might be related to defects in the optical coatings and in the fused silica plates or to an inhomogeneity of the magnetic field.

Theoretically, the demonstrated concept can be applied at various emission wavelengths. The necessary propagation length or magnetic field even reduces when going to the ultraviolet wavelength range due to the increasing Verdet constant. In the UV-vis range, fused silica, quartz, and $MgF_2$ might be promising candidates for the application in a Faraday isolator when used in combination with a simple Herriott cell, although the intrinsic birefringence of the latter two materials might complicate their application.

When going towards a wavelength of 1 μm, further scaling becomes necessary. Employing a stronger magnetic field with permanent magnets like in state-of-the-art Faraday isolators will provide much stronger fields of about 2 – 3 T [29–32], which should be sufficient to reach the desired rotation at a wavelength of 1 μm with fused silica. Further scaling potential lies in the number of passes. For a Herriott cell, mirrors with a larger diameter would provide space for more reflections and, therefore, more passes. Or shorter distances between the cell mirrors would decrease the laser mode diameter on the mirrors and would also increase the possible number of passes. By this means, it appears to be within reach to realize a Faraday rotator with fused silica or even crystalline quartz for high-power lasers at a wavelength of 1 μm. At high powers, the thermal lens appearing within the Faraday medium can be potentially compensated by adjusting the distance of the cell mirrors.

Going further into the mid-IR region, Verdet constants are dropping even more. Here again, the multipass approach employing a Herriott cell could be used to enable Faraday isolators at wavelengths of several μm with materials with usually insufficient Verdet constants. E.g., ZnSe, which transmits wavelengths up to 20 μm, has a Verdet constant of about only 8 rad/(T·m) at a wavelength of 2 μm [33]. A Faraday isolator based on ZnSe proved already to be technically challenging, involving strong magnetic fields and the cascading of two Faraday rotators [33]. Employing the multipass approach shown here with a Herriott cell could reduce the demand for a high magnetic field and the length of the ZnSe rod. Moreover, this approach allows the use of ZnSe as a Faraday medium at much longer wavelengths where the Verdet constant is even lower.

For high-power lasers, it might be advantageous to switch from fused silica to high-quality crystalline quartz to reduce the absorption and thermal effects further. As this material exhibits birefringence, special care has to be taken on its effect on the polarization. To reduce the effect of the birefringence, a plate of c-cut quartz could be used where the c-axis is oriented parallel to the beam propagation direction. The polarization rotation by the optical activity of the material would compensate itself due to the forward and backward propagation in the cell. However, inside the Herriott cell, the angle of incidence on the plate always deviates from perpendicular incidence, leading to a weak birefringence perceived by the beam. The angle can be reduced by considering long cells to minimize the effect of birefringence. Another approach would be to employ a pair of a-cut plates with orthogonal c-axes to compensate for the birefringence. The same considerations apply to other birefringent materials like $MgF_2$. YAG and sapphire might also be interesting candidates as Faraday media due to their compatibility with high-power laser radiation.

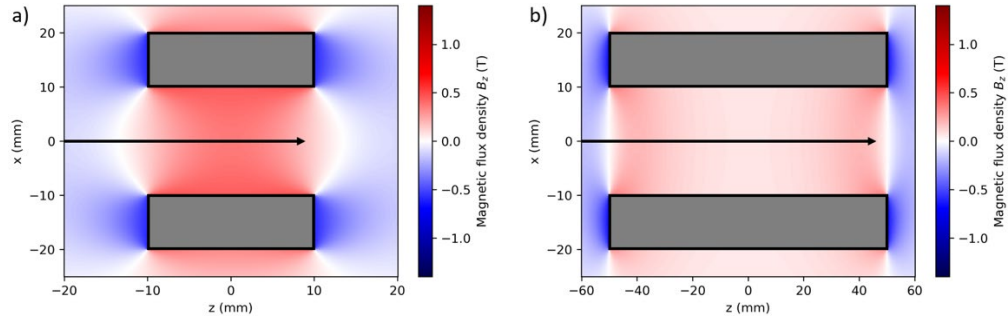

**Figure 4.** Comparison of the magnetic field in z-direction in two ring magnets (grey) with the lengths of 20 mm (a) and 100 mm (b) based on a simulation with magpylib [27]. In both cases, a homogeneous remanent magnetization inside the magnet of 1.4 T is assumed. The beam propagation direction along the z-axis is indicated by a black arrow. The peak magnetic field on the z-axis decreases from 0.36 to 0.19 T, whereas the B-field integrated over the plotted area of the z-axis increases from 5 to 10 mT·m by a factor of two while the length of the magnet increases by a factor of five.

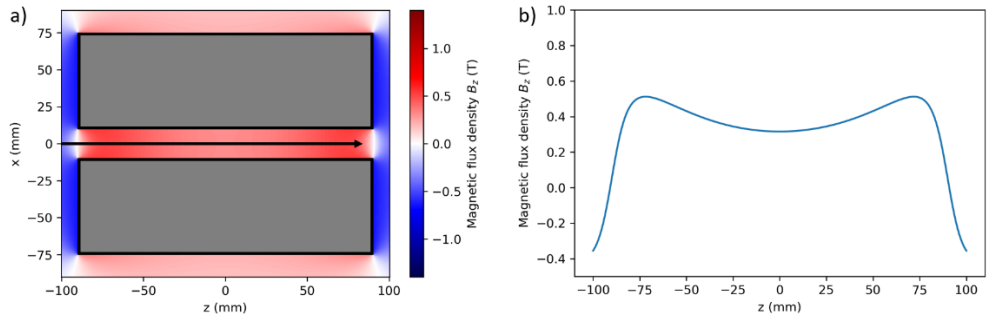

**Figure 5.** (a) Simulation of the z-component of the magnetic field inside a ring magnet (grey) with an outer diameter of 150 mm, an inner diameter of 20 mm, and a length of 180 mm based on magpylib [27]. A homogeneous remanent magnetization inside the magnet of 1.4 T is assumed. The beam propagation direction along the z-axis is indicated by a black arrow. The z-component of the magnetic field integrated over the z-axis yields a value of 66 mT·m. (b) Simulated z-component of the magnetic field on the z-axis going through the center of the ring magnet.

When increasing the average power passing through a Faraday isolator, eventually, absorption and thermally induced birefringence will limit the isolation performance. The use of gases as Faraday media might be a potential way to circumvent the typical limitations arising in solids. Even though very low Verdet constants become more challenging here, the multiple passes from a Herriott cell could help to overcome this obstacle. As a gas, Xenon still has a relatively high Verdet constant of $18 \cdot 10^{-3}$ rad/(T·m·bar) at 500 nm (approx. $4 \cdot 10^{-3}$ rad/(T·m·bar) at 1 µm) [10,13]. To increase the effect, the whole isolator can be built up in a pressurized gas cell. Considering a pressure of 20 bar, this would lead to a Verdet constant of about 0.35 rad/(T·m), still an order of magnitude lower than in fused silica. This deficiency needs to be compensated with a longer propagation length inside the medium. Extending the length of the gas cell leads to the challenge of providing a magnetic field with sufficient flux density over the entire length of the gas cell. It turns out that simply extending the length of a given permanent ring magnet will result in a diminishing magnetic field within the magnet [Fig. 4]. The underlying reason here is the reversion of the magnetic field direction for a magnetic dipole, leading to destructive field components inside the ring magnet. The field can be restored by increasing the outer diameter of the ring magnet. The field of a hypothetical magnet still within practical dimensions is depicted in Fig. 5. Here, the relevant parameter for the Faraday rotation is the field integrated over the beam path. For this magnet, a value of 66 mT·m was calculated. With 20 bar of Xenon and 35 passes, a theoretical rotation angle of 45° results for a wavelength of 500 nm. In the UV range, even smaller magnets with weaker fields can be applied due to the increasing Verdet constant. These considerations point to the general feasibility of a Faraday rotator based on gases as a Faraday medium.

## 5. Conclusion

In summary, we presented a method to enable the use of materials with weak Faraday effect, i.e., low Verdet constants, as a Faraday medium in a Faraday isolator. Guiding the incident radiation multiple times through the Faraday medium in a Herriot cell leads to a summation of the Faraday rotation angles from all passes due to the nonreciprocal nature of the Faraday effect. The desired rotation angle of 45° can be reached with materials otherwise non-applicable in common Faraday isolator schemes. Thereby, materials, e.g., fused silica, become of interest for potential applications in ultraviolet, mid-infrared, and high-power Faraday isolators. Even gases could be principally used, which might turn out beneficial in avoiding limiting effects normally appearing in solids, like thermal lensing or stress-induced birefringence. In a proof-of-principle experiment, we demonstrated the applicability of the concept for fused silica. The polarization of a 532 nm laser has been rotated by an angle of 45° in 12.7 mm of fused silica with the help of a Herriott-type multipass cell.

**Funding.** The research was completely financed by our university budget. No external funding.

**Acknowledgments.** We would like to thank Prof. Dr. Detlef Kip and Dr. Kore Hasse of the Helmut Schmidt University for providing us with their Coherent Verdi laser for the duration of the experiments and for allowing us to use their laboratory.

**Disclosures.** The authors declare no conflicts of interest.

**Data availability.** Data underlying the results presented in this paper are available on request.